\documentclass [journal,onecolumn,11pt]{IEEEtran}
\usepackage{amsfonts,amsmath,amssymb}
\usepackage{indentfirst, setspace}
\usepackage{url,float}
\usepackage{color}
\usepackage[a4paper,ignoreall]{geometry}
\usepackage{longtable}
\usepackage{graphicx}
\usepackage[a4paper,ignoreall]{geometry}
\usepackage{multicol}
\usepackage{stfloats}
\usepackage{enumerate}
\usepackage{cite}
\usepackage{amsthm}
\usepackage{mathrsfs}
\usepackage{multirow}
\usepackage{amssymb}
\usepackage[overload]{empheq}
\usepackage[square, comma, sort&compress, numbers]{natbib}

\def\qu#1 {\fbox {\footnote {\ }}\ \footnotetext { From Qu: {\color{red}#1}}}
\def\hqu#1 {}
\def\kq#1 {\fbox {\footnote {\ }}\ \footnotetext { From KangQuan: {\color{blue}#1}}}
\def\hkq#1 {}

\newtheorem{Th}{Theorem}[section]

\newtheorem{Lemma}[Th]{Lemma}

\newcommand{\tr}{{\rm Tr}}

\newcommand{\gf}{{\mathbb F}}

\newcommand{\Z}{\mathbb {Z}}

\makeatletter
\newcommand{\figcaption}{\def\@captype{figure}\caption}
\newcommand{\tabcaption}{\def\@captype{table}\caption}
\makeatother

\begin{document}
	\title{On a conjecture about a class of permutation quadrinomials}
	\author{{Kangquan Li,  Longjiang Qu, Chao Li and Hao Chen }
	\thanks{Kangquan Li, Longjiang Qu and Chao Li  are with the College of Liberal Arts and Sciences,
		National University of Defense Technology, Changsha, 410073, China.
		Longjiang Qu is also with the State Key Laboratory of Cryptology, Beijing, 100878, China. Hao Chen is with the College of Information Science and Technology/College of Cyber Security, Jinan University, Guangzhou, Guangdong, 510632, China.
		E-mail: likangquan11@nudt.edu.cn, ljqu\_happy@hotmail.com, lichao\_nudt@sina.com and haochen@jnu.edu.cn.
				 This work is supported by the Nature Science Foundation of China (NSFC) under Grant 11531002, 61722213,  61572026,   National Key R$\&$D Program of China (No.2017YFB0802000),  and the Open Foundation of State Key Laboratory of Cryptology.
}}
	\maketitle{}

\begin{abstract}
	Very recently, Tu et al. presented a sufficient condition about $(a_1,a_2,a_3)$, see Theorem \ref{sufficient}, such that $f(x) = x^{3\cdot 2^m} + a_1 x^{2^{m+1}+1}+ a_2 x^{2^m+2} + a_3 x^3$ is a class of permutation polynomials over $\gf_{2^{n}}$ with $n=2m$ and $m$ odd. In this present paper, we prove that the sufficient condition is also necessary.
\end{abstract}


\section{Introduction}

Let $\gf_q$ be the finite field with $q$ elements and $\gf_q^{*}$ be the multiplicative group with the nonzero elements in $\gf_q$. A polynomial $f\in\gf_q[x]$ is called a \emph{permutation polynomial} (PP) if the induced mapping $x\to f(x)$ is a permutation of $\gf_q$. The study about PPs over finite fields attracts people's interest for many years due to their wide applications in coding theory, cryptography and combinatorial designs. 

PPs with few terms attract people's interest due to their simple algebraic form and additional extraordinary properties.  Recently, many scholars studied permutation trinomials over $\gf_{2^{n}}$ with $n=2m$ from Niho exponents of the form 
$$f(x) = x^r\left( 1+a_1x^{s(2^m-1)}+a_2x^{t(2^m-1)} \right), $$
where  $a_1,a_2\in\gf_{2^{n}}$ and the integers $s,t$ can be viewed as elements of $\Z/{(2^m+1)\Z}$, see \cite{LQC2017,Bartoli2018,Tu2018,Hou2018,LQLF2018,Tu2018two,li2019new,hou2019}, etc. For more relative results, readers can refer in two recent survey papers \cite{wang2019,li2019survey}. However, up to now, there are only four cases of parameters $(r, s,t)$ that have been determined completely:
\begin{enumerate}[(1)]
\item $(r, s,t) = (1, 1,2)$ \cite{hou2015determination};
\item $(r, s,t) = (1, -1/2, 1/2)$ \cite{Tu2018two};
\item $(r, s,t) = (1, -1,2)$ \cite{Bartoli2018,Hou2018,Tu2018};
\item $(r, s,t)= (1, 1/4, 3/4)$ \cite{hou2019,Tu2018two}.
\end{enumerate}

 Very recently, in \cite{TLZ2019}, the authors presented a class of permutation quadrinomials as follows. Note that for each element $x$ in $\gf_{2^{n}}$ with $n=2m$, we define $\overline{x}=x^{2^m}$. Moreover,  we use  $\tr_1^m(\cdot)$ to denote the absolute trace function from $\gf_{2^{m}}$ to $\gf_{2}$, i.e., for any $x\in\gf_{2^{m}}$,
 $\tr_1^m(x)=x+x^2+\cdots+x^{2^{m-1}}.$

\begin{Th}
	\cite{TLZ2019}
	\label{sufficient}
	Let $n=2m$ for odd $m$ and define 
	$$\Gamma = \left\{ (a_1,a_2,a_3): \theta_2^2=\theta_1\overline{\theta}_3, \theta_1 \neq0, \tr_1^m\left(\frac{\theta_4}{\theta_1}\right) =1, a_1\in\gf_{2^{m}}, a_2, a_3\in\gf_{2^{n}} \right\}, $$
	where 
	\begin{equation}
	\label{theta}
	\theta_1 = 1+a_1^2+a_2\overline{a}_2+a_3\overline{a}_3, \theta_2 = a_1+\overline{a}_2 a_3, \theta_3 = \overline{a}_2 + a_1\overline{a}_3, \theta_4 = a_1^2+a_2\overline{a}_2. 
	\end{equation}
	Then for any $(a_1,a_2,a_3)\in\Gamma$, the quadrinomial
	\begin{eqnarray*}
	f(x) = \overline{x}^3+a_1\overline{x}^2x+a_2x^2\overline{x}+a_3x^3
	\end{eqnarray*}
	is a permutation of $\gf_{2^{n}}$.
\end{Th}

The authors in \cite{TLZ2019} conjectured that the sufficient condition in Theorem  \ref{sufficient} is also necessary. In the present paper, we prove that their conjecture is right. Namely, 

\begin{Th}
	\label{main_theorem}
	Let $n=2m$. Then the quadrinomial
	$$f(x) = \overline{x}^3+a_1\overline{x}^2x+a_2x^2\overline{x}+a_3x^3$$
	is a permutation of $\gf_{2^{n}}$ if and only if $m$ is odd and $(a_1,a_2,a_3)\in\Gamma$, where $\Gamma$ is defined as in Theorem \ref{sufficient}.
\end{Th}

Firstly, as pointed out in \cite{TLZ2019}, the assumption $a_1\in\gf_{2^{m}}$ is reasonable since 
\begin{eqnarray*}
f(\beta x) &=& (\overline{\beta}\overline{x})^3+a_1(\overline{\beta}\overline{x})^2\beta x + a_2\beta^2 x^2 \overline{\beta}\overline{x} + a_3(\beta x)^3 \\
&=& \overline{\beta}^3\left( \overline{x}^3+a_1 (\beta/\overline{\beta}) \overline{x}^2x+a_2(\beta/\overline{\beta})^2x^2\overline{x} +a_3 (\beta/\overline{\beta})^3x^3    \right),
\end{eqnarray*} 
where $\beta\in\gf_{2^{n}}^{*}$ satisfies $\beta^2 a_1=1$ and $a_1 (\beta/\overline{\beta})=\beta^{-1-2^m}\in\gf_{2^{m}}$.
Next,  if  $1+a_1+a_2+a_3=0$, then $f(0)=f(1)=0$ and $f$ is not a permutation. Thus in this paper, we always assume $$1+a_1+a_2+a_3\neq0.$$
In addition, it is clear that $f(x)=x^3h\left(x^{2^m-1}\right)$, where $h(x)=x^3+a_1x^2+a_2x+a_3$. The PPs of the form  {$x^rh\left(x^{(q-1)/d}\right)$} over $\gf_q$ are interesting and have been  paid particular attention. There is a connection between the PPs of this type and certain permutations of the subgroup of order $d$ of $\gf_q^\ast$. 
\begin{Lemma}
	\label{lem1}
	\cite{park2001permutation,wang2007cyclotomic,zieve2009some}
	Pick $d,r > 0$ with $d\mid (q-1)$, and let $h(x)\in\gf_q[x]$. Then $f(x)=x^rh\left(x^{\left.(q-1)\middle/d\right.}\right)$ permutes $\mathbb{F}_q$ if and only if both
	\begin{enumerate}[(1)]
		\item $\gcd(r,\left.(q-1)\middle/d\right.)=1$ and
		\item $g(x)=x^rh(x)^{\left.(q-1)\middle/d\right.}$ permutes $\mu_d$, where $\mu_d=\{x\in\gf_q : x^d=1\}$.
	\end{enumerate}
\end{Lemma}

Since $\gcd(3,2^m-1)=1$ if and only if $m$ is odd, together with Lemma \ref{lem1},   we know that $f(x)$ is a PP over $\gf_{2^{n}}$ if and only if $m$ is odd and  
\begin{eqnarray*}
g(x) &=& x^3h(x)^{2^m-1} \\
&=& \frac{\overline{a}_3x^3+\overline{a}_2x^2+a_1x+1}{x^3+a_1x^2+a_2x+a_3}
\end{eqnarray*}
permutes $\mu_{2^m+1}$. Let $\phi(x) = \frac{x+\omega^2}{x+\omega}$ be a bijection from $\gf_{2^m}$ to $\mu_{2^m+1}\backslash\{1\}$, where $\omega\in\gf_{2^{n}}\backslash\gf_{2^{m}}$ and $\omega^2+\omega+1=0$. Then $g$ permutes $\mu_{2^m+1}$ if and only if $g(\phi(x))$ is a bijection. This happens if and only if 
\begin{equation}
\label{F(x)}
F(x) = g(\phi(x))|_{\gf_{2^m}}: \gf_{2^{m}} \to \mu_{2^m+1}\backslash \{ (1+a_1+a_2+a_3)^{2^m-1} \}
\end{equation}
is a bijection. Let $$L(x,y) = N\left(\frac{F(x)-F(y)}{x-y}\right) =0,$$ where $N\left(\frac{F(x)-F(y)}{x-y}\right)$ denotes the numerator of $\frac{F(x)-F(y)}{x-y}$. Clearly, $F(x)$ is a bijection if and only if $L(x,y)=0$ has no $\gf_{2^m}$-rational points off the line $x=y$. 

Our method to show that the sufficient condition in Theorem  \ref{sufficient} is also necessary is based on the Hasse-Weil bound, which has been applied in the study of PPs, e.g. \cite{Bartoli2018,Hou2018}.  
\begin{Lemma}
	\cite[Hasse-Weil bound]{Houbook2018}
	\label{Hasse-weil}
	Let $L(x,y)$ be a polynomial in $\gf_{q}[x,y]$ of degree $d$ and let $\# V_{\gf_{q}^2}(L)$ be the number of zeros of $L$. If $L$ has an absolutely irreducible component over $\gf_{q}$, then 
	$$ \left| \# V_{\gf_{q}^2}(L)-q \right| \le (d-1)(d-2)q^{1/2} + \frac{1}{2}d(d-1)^2+1. $$
\end{Lemma} 
In addition, in our proof, some relations obtained in \cite{TLZ2019} between $\theta_i$ for $i=1,2,3,4$ are also useful. 
\begin{Lemma}	\label{TLZ} \cite{TLZ2019} For $\theta_i$ ($i=1,2,3,4$) defined as (\ref{theta}), we have
\begin{enumerate}
	\item  $\theta_2\overline{\theta}_2+\theta_3\overline{\theta}_3=\theta_4(\theta_1+\theta_4)$ (note that the relation always holds just from the definition of $\theta_i$, without the assumption $(a_1,a_2,a_3)\in\Gamma$);
	\item if $(a_1,a_2,a_3)\in\Gamma$, we have $\theta_2\theta_3+\overline{\theta}_2\overline{\theta}_3=\frac{\theta_2\overline{\theta}_2(\theta_2+\overline{\theta}_2)}{\theta_1}$;
	\item if $(a_1,a_2,a_3)\in\Gamma$, if there exists an element $\lambda\in\mu_{2^m+1}$ such that $\theta_1+\theta_2\overline{\lambda}+\overline{\theta}_2\lambda=0$, then  $\theta_2\overline{\theta}_2=\theta_1\theta_4$.
\end{enumerate}
\end{Lemma}

In the following, we firstly determine the necessary and sufficient conditions about $(a_1,a_2,a_3)$ for $L(x,y)$ to split completely into absolutely irreducible components not defined over $\gf_{2^m}$, see Theorem \ref{factor_L}, whose proof will be given in the next section.  For convenience, if $L$ can be factorized as a product of four linear factors, we write $L=(1,1,1,1)$; if $L$ can be factorized as a product of two quadratic absolute irreducible factors, we write $L=(2,2)$.

\begin{Th}
	\label{factor_L}
	Let $L(x,y)\neq0$ defined as (\ref{L(x,y)}) and $1+\theta_1+\theta_2+\theta_3\neq0$. Then the factorizations of $L(x,y)$  into absolute irreducible components not defined over $\gf_{2^{m}}$ are characterized as follows:
	\begin{enumerate}[(a)]
		\item $L=(1,1,1,1)$ if and only if $$\theta_1\neq0,  \theta_1\neq\theta_3+\overline{\theta}_3, \theta_2^2=\theta_1\overline{\theta}_3, \theta_4^2 = \theta_1^2+\theta_3\overline{\theta}_3, \tr_1^{m}\left(\frac{\theta_4}{\theta_1}\right)=1;$$
		\item $L=(2,2)$ if and only if $$\theta_1\neq0,  \theta_1\neq\theta_3+\overline{\theta}_3, \theta_2^2=\theta_1\overline{\theta}_3, \tr_1^m\left(\frac{\theta_4}{\theta_1}\right)=1;$$
		\item $L=(1,1)$ if and only if $$ \theta_1=\theta_2+\overline{\theta}_2\neq0, \theta_3=\theta_2+\gamma, \theta_4=\gamma+\theta_2+\overline{\theta}_2 ~~\text{and}~~ \gamma=\frac{\theta_2^2+\overline{\theta}_2^2+\theta_2\overline{\theta}_2}{\theta_2+\overline{\theta}_2}. $$
	\end{enumerate}
\end{Th}

In the last of this section, we give the proof of Theorem \ref{main_theorem}. 

\begin{proof}
	Firstly, according to the item (1) of Lemma \ref{lem1}, we know that if $f$ is a PP, then $m$ is odd since $\gcd(3,2^m-1)=1$ if and only if $m$ is odd.
	Let $$ \Gamma_1 := \{ (a_1,a_2,a_3)\in\Gamma: \theta_1\neq\theta_3+\overline{\theta}_3 \} ~~\text{and}~~ \Gamma_2 := \{ (a_1,a_2,a_3)\in\Gamma: \theta_1=\theta_3+\overline{\theta}_3 \}. $$
	Then,  it is clear that the condition of (a) in Theorem \ref{factor_L} belongs to $\Gamma_1$ and that of (b) is indeed $\Gamma_1$. In the following, we show that the condition of (c) in Theorem \ref{factor_L} is  $\Gamma_2$. 
	Recall that 	$$\Gamma = \left\{ (a_1,a_2,a_3): \theta_2^2=\theta_1\overline{\theta}_3, \theta_1 \neq0, \tr_1^m\left(\frac{\theta_4}{\theta_1}\right) =1, a_1\in\gf_{2^{m}}, a_2, a_3\in\gf_{2^{n}} \right\}. $$
	Then for $(a_1,a_2,a_3)\in\Gamma_2,$ i.e., $\theta_1=\theta_3+\overline{\theta}_3$, we have $\theta_2^2=\theta_1\overline{\theta}_3=(\theta_3+\overline{\theta}_3)\overline{\theta}_3$ and thus $\overline{\theta}_2^2=(\theta_3+\overline{\theta}_3){\theta}_3$. Adding the above two equations, we obtain $\theta_2+\overline{\theta}_2=\theta_3+\overline{\theta}_3$. Let $\gamma=\theta_2+\theta_3$. Then $\gamma\in\gf_{2^{m}}$. In addition, from Lemma \ref{TLZ}, we have $\theta_2\theta_3+\overline{\theta}_2\overline{\theta}_3=\frac{\theta_2\overline{\theta}_2(\theta_2+\overline{\theta}_2)}{\theta_1}$ and thus $\theta_2\theta_3+\overline{\theta}_2\overline{\theta}_3=\theta_2\overline{\theta}_2$. Plugging $\theta_3=\theta_2+\gamma$ and $\overline{\theta}_3=\overline{\theta}_2+\gamma$ into the above equation, we get $$\gamma=\frac{\theta_2^2+\overline{\theta}_2^2+\theta_2\overline{\theta}_2}{\theta_2+\overline{\theta}_2}.$$
	Furthermore, from Lemma \ref{TLZ}, we have $\theta_2\overline{\theta}_2=\theta_1\theta_4$. Thus $$\theta_4=\frac{\theta_2\overline{\theta}_2}{\theta_2+\overline{\theta}_2}=\gamma+\theta_2+\overline{\theta}_2.$$ 
	Therefore, the condition of (c) in Theorem \ref{factor_L} is indeed $\Gamma_2$. 
	
	All in all, $L$ can split completely into absolute irreducible components not defined over $\gf_{2^{m}}$ if and only if $(a_1,a_2,a_3)\in\Gamma$. When $(a_1,a_2,a_3)\not\in\Gamma$, $L$ with degree $4$ has an absolutely irreducible factor over $\gf_{2^{m}}$ and thus according to Lemma \ref{Hasse-weil}, we have 
	\begin{eqnarray*}
		\# V_{\gf_{2^m}^2}(L) &\ge& 2^m-(d-1)(d-2)2^{m/2}-\frac{1}{2}d(d-1)^2-1 \\
		&=& 2^m-6\cdot 2^{m/2} - 19.
	\end{eqnarray*}
	
	Let $\lambda\in\mathbb{R}$ denote the larger solution of $x^2-6 x-21=0.$ Then $\lambda< 9$ and thus
	$\# V_{\gf_{2^m}^2}(L) > 2 $ when $m\ge4$. By (\ref{L(x,y)}), $L(x,x)= \ell_{22}x^4+\ell_{11}x^2+\ell_{00}$, which has at most two zeros in $\gf_{2^{m}}$. Hence $L(x,y)$ has at least a zero $(x,y)\in\gf_{2^{m}}^2$ with $x\neq y$ when $m\ge 4$. Consequently, when $(a_1,a_2,a_3)\not\in\Gamma$ with $m\ge 4$, $f$ is not a permutation over $\gf_{2^{n}}$. For the case $m<4$, the same conclusion can be obtained by the MAGMA. 
	
	Together with the sufficient proof in \cite{TLZ2019}, the proof of Theorem \ref{main_theorem} has been finished completely. 
\end{proof}

Therefore, in the following we only suffice to provide the proof of Theorem \ref{factor_L}, which will be given in the next section. Through the paper, the algebraic closure of $\gf_{2^{m}}$ is denoted by $\overline{\gf}_{2^m}$.

\section{The proof of Theorem \ref{factor_L}}
In this section, we prove Theorem \ref{factor_L}.
After direct computation, we have 
\begin{eqnarray*}
F(x) &=& g\left( \phi(x) \right) \\ &=&\frac{\overline{a}_3\left(\frac{x+\omega^2}{x+\omega}\right)^3+\overline{a}_2\left(\frac{x+\omega^2}{x+\omega}\right)^2+a_1\left(\frac{x+\omega^2}{x+\omega}\right)+1}{\left(\frac{x+\omega^2}{x+\omega}\right)^3+a_1\left(\frac{x+\omega^2}{x+\omega}\right)^2+a_2\left(\frac{x+\omega^2}{x+\omega}\right)+a_3} \\
&=& \frac{\epsilon_1x^3+\epsilon_2x^2+\epsilon_3x+\epsilon_4}{\tau_1x^3+\tau_2x^2+\tau_3x+\tau_4},
\end{eqnarray*}

where 
	\begin{subequations} 	
	\renewcommand\theequation{\theparentequation.\arabic{equation}}     
	\begin{empheq}[left={\empheqlbrace\,}]{align}
~~	\epsilon_1 &= a_1+\overline{a}_2+\overline{a}_3+1   \notag \\ 
~~	 \epsilon_2 &= \omega^2 a_1+\omega \overline{a}_2+ \omega^2 \overline{a}_3 +\omega     \notag \\
~~	 \epsilon_3 &= \omega^2 a_1 + \omega \overline{a}_2+\omega \overline{a}_3 + \omega^2 \notag\\
~~	 \epsilon_4 &= \omega a_1+\omega^2 \overline{a}_2+\overline{a}_3+1, \notag
	\end{empheq}
\end{subequations}
and 
	\begin{subequations} 	
	\renewcommand\theequation{\theparentequation.\arabic{equation}}     
	\begin{empheq}[left={\empheqlbrace\,}]{align}
~~	\tau_1 &= a_1+{a}_2+{a}_3+1   \notag \\ 
~~	\tau_2 &= \omega a_1+\omega^2 {a}_2+ \omega {a}_3 +\omega^2     \notag \\
~~	\tau_3 &= \omega a_1 + \omega^2 {a}_2+\omega^2 {a}_3 + \omega \notag\\
~~	\tau_4 &= \omega^2 a_1+\omega {a}_2+ {a}_3+1. \notag
	\end{empheq}
\end{subequations}
Therefore, we have 
\begin{equation}
\label{L(x,y)}
L(x,y) = N\left( \frac{F(x)-F(y)}{x-y} \right)   = \ell_{22} x^2y^2+\ell_{21}x^2y + \ell_{12} xy^2+\ell_{20} x^2+ \ell_{11}xy + \ell_{02} y^2+ \ell_{10} x + \ell_{01} y + \ell_{00},
\end{equation}
where 
\begin{eqnarray*}
\ell_{22} &=& a_1^2+a_1a_3+a_1\overline{a}_3+a_2\overline{a}_2+a_2+a_3\overline{a}_3+\overline{a}_2+1\\
&=& \theta_1+\theta_3+\overline{\theta}_3,
\end{eqnarray*}
\begin{eqnarray*}
\ell_{21} &=& a_1^2+a_2\overline{a}_2 +a_2\overline{a}_3 + a_3\overline{a}_2 + a_3\overline{a}_3+1 \\
&=& \theta_1+\theta_2+\overline{\theta}_2, 
\end{eqnarray*}
\begin{eqnarray*}
\ell_{20} &=& a_1^2+\omega^2a_1a_3+\omega a_1\overline{a}_3+a_1+a_2\overline{a}_2 + \omega^2a_2\overline{a}_3+\omega^2 a_2 +\omega a_3\overline{a}_2 + \omega \overline{a}_2 \\
&=& \theta_4 + \omega \theta_3 + \omega^2 \overline{\theta}_3 + \omega\theta_2 + \omega^2 \overline{\theta}_2,
\end{eqnarray*}
$$\ell_{11}=a_1^2+a_2\overline{a}_2+a_3\overline{a}_3+1 = \theta_1,$$
\begin{eqnarray*}
\ell_{10} &=& a_1^2+a_1+a_2\overline{a}_2+\omega a_2\overline{a}_3 + \omega^2  a_3 \overline{a}_2 + a_3\overline{a}_3 +1\\
&=& \theta_1+\omega^2\theta_2 + \omega\overline{\theta}_2,
\end{eqnarray*}
\begin{eqnarray*}
\ell_{00} &=& a_1^2+\omega a_1a_3 + \omega^2 a_1\overline{a}_3 + a_2\overline{a}_2 + \omega a_2+a_3\overline{a}_3+\omega^2\overline{a}_2+1\\
&=&\theta_1 + \omega^2 \theta_3 + \omega \overline{\theta}_3,
\end{eqnarray*}
and $\ell_{12}=\ell_{21}$, $\ell_{02} = \ell_{20}$, $\ell_{01}=\ell_{10}$, $\theta_i$ for $i=1,2,3,4$ are defined as (\ref{theta}).

In the following, we determine the necessary and sufficient conditions for $L(x,y)$ to split completely into absolutely irreducible components not defined over $\gf_{2^m}$.  The factorization can be divided into two cases:  $\ell_{22}\neq0$ and  $\ell_{22}=0$.

	{\bfseries Case 1: $\ell_{22}\neq0$.} Namely, $\theta_1\neq\theta_3+\overline{\theta}_3$.

In the case, $\deg L =4$ and it is clear that the morphisms $(x,y)\to (y,x)$ and $(x,y) \to (\overline{x},\overline{y})$ fix $L=0$ and therefore they act on its components, which means that if $x+a$ with $a\in\overline{\gf}_{2^m}\backslash\gf_{2^{m}}$ is a component  of $L$, then $y+a$ and $x+\overline{a}$ are also the components of $L$ directly.  Moreover, if $L(x,y)$ can split completely into absolute irreducible components not defined over $\gf_{2^m}$, then the only possibilities are $(1,1,1,1)$ and $(2,2)$. The reason is as follows. If $x+a$ with $a\in\overline{\gf}_{2^m}\backslash\gf_{2^{m}}$ is a component {not defined over $\gf_{2^{m}}$} of $L$, then $y+a$, $x+\overline{a}$ and $y+\overline{a}$ are also the components of $L$, i.e., $L=(1,1,1,1)$. If $L$ has a component {not defined over $\gf_{2^{m}}$} with degree $2$, denoted by $L_1$, then $\overline{L}_1$ with degree $2$ is also a component of $L$, where $\overline{L}_1$ denotes the polynomial from raising all the coefficients of $L_1$ into their $2^m$-th power respectively. Obviously, $L_1\neq \overline{L}_1$. Thus, $L=(2,2)$, i.e., $L$ can not be $(2,1,1)$. In addition, the impossibility of $L=(3,1)$ is trivial.

{\bfseries Subcase 1.1: } $L= (1,1,1,1)$. In the subcase, there must exist some $a\in\overline{\gf}_{2^m}\backslash\gf_{2^{m}}$ such that 
\begin{eqnarray*}
	L &=& \ell_{22}(x+a)(x+\overline{a})(y+a)(y+\overline{a}) \\
	&=& \ell_{22}\left(x^2y^2+ (a+\overline{a})x^2y + a\overline{a}x^2 + (a+\overline{a}) xy^2 + (a^2+\overline{a}^2) xy \right. \\
	&& \left.+ (a^2\overline{a}+a\overline{a}^2)x + a\overline{a} y^2+(a^2\overline{a}+a\overline{a}^2) y+a^2\overline{a}^2 \right).
\end{eqnarray*} 
Comparing the coefficients of the above expression and (\ref{L(x,y)}), we have 
\begin{subequations} 	
	\renewcommand\theequation{\theparentequation.\arabic{equation}}     
	\begin{empheq}[left={\empheqlbrace\,}]{align}
	~~	\ell_{21} &=  (a+\overline{a})\ell_{22}  \label{l_1} \\ 
	~~	\ell_{20} &= a\overline{a}\ell_{22}     \label{l_2} \\
	~~	\ell_{11} &= (a^2+\overline{a}^2)\ell_{22} \label{l_3} \\
	~~	\ell_{10} &= (a^2\overline{a}+a\overline{a}^2)\ell_{22} \label{l_4} \\
	~~	\ell_{00} &= a^2\overline{a}^2\ell_{22}. \label{l_5}
	\end{empheq}
\end{subequations}
Computing $(\ref{l_1})^2+ \ell_{22}\times (\ref{l_3}) $, $ (\ref{l_2})^2+ \ell_{22}\times (\ref{l_5}) $ and $(\ref{l_1})\times (\ref{l_2})+ \ell_{22} \times (\ref{l_4})$ respectively, we obtain 
\begin{subequations} 	
	\renewcommand\theequation{\theparentequation.\arabic{equation}}     
	\begin{empheq}[left={\empheqlbrace\,}]{align}
	~~ &	\ell_{21}^2 =  \ell_{11}\ell_{22} \notag   \\ 
	~~&	\ell_{20}^2 = \ell_{00}\ell_{22}    \notag  \\
	~~&	\ell_{21}\ell_{20} = \ell_{10}\ell_{22}, \notag
	\end{empheq}
\end{subequations}
namely,
\begin{subequations} 
	\small	
	\renewcommand\theequation{\theparentequation.\arabic{equation}}     
	\begin{empheq}[left={\empheqlbrace\,}]{align}
	& \theta_1\theta_3 + \theta_1 \overline{\theta}_3+\theta_2^2+\overline{\theta}_2^2 =0  \label{the_1}   \\ 
	& \theta_1^2 + \omega \theta_1 \theta_3 + \omega^2 \theta_1\overline{\theta}_3 + \omega^2\theta_2^2+ \omega \overline{\theta}_2^2 + \theta_3\overline{\theta}_3+\theta_4^2=0  \label{the_2}  \\
	&	\theta_1^2 + \theta_1\theta_2 + \omega^2\theta_1\theta_3+\theta_1\theta_4+\theta_1\overline{\theta}_2+\omega\theta_1\overline{\theta}_3+\omega\theta_2^2+\theta_2\theta_3+\theta_2\theta_4+\theta_2\overline{\theta}_2+\theta_4\overline{\theta}_2+\omega^2\overline{\theta}_2^2+\overline{\theta}_2\overline{\theta}_3=0. \label{the_3}
	\end{empheq}
\end{subequations}
Computing $(\ref{the_2})\times (\theta_1+\theta_2+\overline{\theta}_2)^2+(\ref{the_3})^2$, we get 
\begin{equation*}
(\theta_1+\theta_3+\overline{\theta}_3)(\omega\theta_1^2\theta_3+\omega^2\theta_1^2\overline{\theta}_3+\omega^2\theta_1\theta_2^2+\omega\theta_1\overline{\theta}_2^2+\theta_2^2\theta_3+\overline{\theta}_2^2\overline{\theta}_3)=0.
\end{equation*}
Since $\ell_{22}=\theta_1+\theta_3+\overline{\theta}_3\neq0$ in the case, we have 
\begin{equation}
\label{the_4}
\omega\theta_1^2\theta_3+\omega^2\theta_1^2\overline{\theta}_3+\omega^2\theta_1\theta_2^2+\omega\theta_1\overline{\theta}_2^2+\theta_2^2\theta_3+\overline{\theta}_2^2\overline{\theta}_3=0.
\end{equation}
Moreover, from Eq. (\ref{the_1}), we can assume that $\theta_2^2+\theta_1 \overline{\theta}_3 = \gamma\in\gf_{2^{m}}$. Then $\theta_2^2=\gamma+\theta_1 \overline{\theta}_3$ and $\overline{\theta}_2^2= \gamma+\theta_1 {\theta}_3$. Plugging them into Eq. (\ref{the_4}), we have
\begin{eqnarray*}
	&&\omega\theta_1^2\theta_3+\omega^2\theta_1^2\overline{\theta}_3+\omega^2\theta_1\theta_2^2+\omega\theta_1\overline{\theta}_2^2+\theta_2^2\theta_3+\overline{\theta}_2^2\overline{\theta}_3 \\
	&=& \omega\theta_1^2\theta_3+\omega^2\theta_1^2\overline{\theta}_3+\omega^2\theta_1\left(\gamma+\theta_1 \overline{\theta}_3\right)+\omega\theta_1\left(\gamma+\theta_1 {\theta}_3\right)+\left( \gamma+\theta_1 \overline{\theta}_3  \right)\theta_3+\left(\gamma+\theta_1 {\theta}_3\right)\overline{\theta}_3\\
	&=& \gamma (\theta_1+\theta_3+\overline{\theta}_3) =0,
\end{eqnarray*}
which means $\gamma=\theta_2^2+\theta_1 \overline{\theta}_3=0$, also thanks to $\ell_{22}=\theta_1+\theta_3+\overline{\theta}_3\neq0$ in the case. Furthermore, $a+\overline{a}=\frac{\ell_{21}}{\ell_{22}
}$ and $a\overline{a}=\frac{\ell_{20}}{\ell_{22}}$ and thus $a,\overline{a}$ are solutions of 
$$x^2+\frac{\ell_{21}}{\ell_{22}}x + \frac{\ell_{20}}{\ell_{22}} =0,$$
which does not have solutions in $\gf_{2^{m}}$ if and only if $$\ell_{21}^2 = \theta_1^2+\theta_2^2+\overline{\theta}_2^2  = \theta_1 (\theta_1+\theta_3+\overline{\theta}_3)\neq0,$$
i.e., $\theta_1\neq0$  and $\tr_1^m\left( \frac{\ell_{20}}{\ell_{22}}  \cdot \frac{\ell_{22}^2}{\ell_{21}^2}  \right) = \tr_1^m\left(\frac{\ell_{20}\ell_{22}}{\ell_{21}^2}\right)=1$. Moreover,
\begin{eqnarray*}
	&& \tr_1^m\left(\frac{\ell_{20}\ell_{22}}{\ell_{21}^2}\right) \\ 
	&=& \tr_1^m\left( \frac{(\theta_1+\theta_3+\overline{\theta}_3)(\theta_4 + \omega \theta_3 + \omega^2 \overline{\theta}_3 + \omega\theta_2 + \omega^2 \overline{\theta}_2)  }{\theta_1^2+\theta_2^2+\overline{\theta}_2^2} \right)\\
	&=&\tr_1^m\left( \frac{(\theta_1+\theta_3+\overline{\theta}_3)(\theta_4 + \omega \theta_3 + \omega^2 \overline{\theta}_3 + \omega\theta_2 + \omega^2 \overline{\theta}_2)  }{\theta_1^2+\theta_1\theta_3+\theta_1\overline{\theta}_3} \right)\\
	&=&\tr_1^m\left( \frac{\theta_4 + \omega \theta_3 + \omega^2 \overline{\theta}_3 + \omega\theta_2 + \omega^2 \overline{\theta}_2 }{\theta_1}\right) \\
	&=& \tr_1^m\left(\frac{\theta_4}{\theta_1}\right)	
\end{eqnarray*}
and thus $\tr_1^m\left(\frac{\theta_4}{\theta_1}\right)=1$. 
Furthermore, when $\theta_2^2=\theta_1\overline{\theta}_3$, it is easy to show that Eq. (\ref{the_2}) and Eq. (\ref{the_3}) holds if and only if $\theta_4^2=\theta_1^2+\theta_3\overline{\theta}_3$.  Thus, $L(x,y)$ can split completely into absolute irreducible components not defined over $\gf_{2^m}$ with $(1,1,1,1)$ if and only if $$\theta_1\neq0, \theta_1\neq\theta_3+\overline{\theta}_3, \theta_2^2=\theta_1\overline{\theta}_3, \theta_4^2 = \theta_1^2+\theta_3\overline{\theta}_3, \tr_1^{m}\left(\frac{\theta_4}{\theta_1}\right)=1.$$

{\bfseries  Subcase 1.2:}  $L= (2,2)$. In the subcase, we firstly consider the possibilities about the factorization of $L$. (i) If $x^2+ay^2+bxy+cx+dy+e$ is a component of $L$, then clearly, $ y^2+ax^2+bxy+cy+dx+e $ is  the other component of $L$. Since the coefficients of $x^4$ and $x^3y$ in $L$ are $0$, we have $a=b=0$. (ii) If $xy+ax+ay+b$ is a component of $L$, then $xy+\overline{a}x+\overline{a}y+\overline{b}$ is the other component of $L$. (iii) If $xy+ax+by+c$ with $b\neq a$ is a component of $L$, then $xy+ay+bx+c$ is the other component of $L$. Moreover, $xy+\overline{a}x+\overline{b}y+\overline{c}$ is also a component of $L$ and thus $xy+\overline{a}x+\overline{b}y+\overline{c}=xy+ay+bx+c$. Furthermore, we have $b=\overline{a}$ and $c\in\gf_{2^{m}}$.

Hence there are three possibilities about  the factorization of $L$:
\begin{enumerate}[(i)]
	\item  $$L=\ell_{22}(x^2+ax+by+c)(y^2+ay+bx+c),$$
	\item  $$L=\ell_{22}(xy+ax+ay+b)(xy+\overline{a}x+\overline{a}y+\overline{b}),$$
	\item $$L=\ell_{22}(xy+ax+\overline{a}y+b)(xy+\overline{a}x+{a}y+b).$$
\end{enumerate}

As for (i), after comparing the coefficients (mainly $x^3$ and $y^3$) of the hypothetic expression of $L$ and (\ref{L(x,y)}), we obtain $b=0$ directly and thus the possibility becomes the Subcase 1.1. 

As for (ii), there exist some $a,b\in\overline{\gf}_{2^m}$ such that
\begin{eqnarray*}
	L &=& \ell_{22}(xy+ax+ay+b)(xy+\overline{a}x+\overline{a}y+\overline{b}) \\
	&=& \ell_{22}\left( x^2y^2 + (\overline{a}+a)x^2y+  (\overline{a}+a)xy^2 + a\overline{a} x^2 + (\overline{b} + b) xy \right. \\
	& & \left.+ a\overline{a} y^2 + (a\overline{b} + \overline{a} b)x + (a\overline{b}+\overline{a}b) y + b\overline{b} \right).
\end{eqnarray*}
Comparing the coefficients of the above expression and (\ref{L(x,y)}), we have 

\begin{subequations} 	
	\renewcommand\theequation{\theparentequation.\arabic{equation}}     
	\begin{empheq}[left={\empheqlbrace\,}]{align}
	~~	\ell_{21} &=  (a+\overline{a})\ell_{22}  \label{2_l_1} \\ 
	~~	\ell_{20} &= a\overline{a}\ell_{22}     \label{2_l_2} \\
	~~	\ell_{11} &= (b+\overline{b})\ell_{22} \label{2_l_3} \\
	~~	\ell_{10} &= (a\overline{b}+ \overline{a} b)\ell_{22} \label{2_l_4} \\
	~~	\ell_{00} &= b\overline{b}\ell_{22}. \label{2_l_5}
	\end{empheq}
\end{subequations}

When $\ell_{21}\neq0$ and $\ell_{11}\neq0$, i.e., $a,b\in \gf_{2^{n}}\backslash \gf_{2^{m}}$, from (\ref{2_l_1}) and (\ref{2_l_2}), we know that $a$ and $\overline{a}$ are solutions of $x^2+\frac{\ell_{21}}{\ell_{22}}x+\frac{\ell_{20}}{\ell_{22}}=0$, which does not have solutions in $\gf_{2^{m}}$ if and only if $\tr_1^m\left( \frac{\ell_{20}\ell_{22}}{\ell_{21}^2} \right)=1$. Thus we can assume that $\frac{\ell_{20}\ell_{22}}{\ell_{21}^2}=1+\gamma+\gamma^2$, where $\gamma\in\gf_{2^{m}}$ and in the case, we have $$a=\frac{\ell_{21}}{\ell_{22}}(\omega+\gamma) ~~\text{and}~~ \overline{a}=\frac{\ell_{21}}{\ell_{22}}(\omega^2+\gamma).$$ Also, the similar discussion about $b$ and $\overline{b}$ holds. We have $\tr_1^m\left(\frac{\ell_{00}\ell_{22}}{\ell_{11}^2} \right)=1$ and assume that $\frac{\ell_{00}\ell_{22}}{\ell_{11}^2}=1+\eta+\eta^2,$ where $\eta\in\gf_{2^{m}}$. Moreover, $$b=\frac{\ell_{11}}{\ell_{22}}(\omega+\eta) ~~\text{and}~~ \overline{b}=\frac{\ell_{11}}{\ell_{22}}(\omega^2+\eta).$$ Furthermore, together with (\ref{2_l_4}) and the expressions of $a,\overline{a},b,\overline{b}$, we obtain $$\frac{\ell_{10}\ell_{22}}{\ell_{21}\ell_{11}}=\gamma+\eta.$$
After direct computing, we get 
$$ \frac{\ell_{00}\ell_{22}}{\ell_{11}^2}+\frac{\ell_{20}\ell_{22}}{\ell_{21}^2}=\frac{\ell_{10}\ell_{22}}{\ell_{21}\ell_{11}}+\frac{\ell_{10}^2\ell_{22}^2}{\ell_{21}^2\ell_{11}^2},  $$
i.e., 
$$\ell_{00}\ell_{21}^2+\ell_{20}\ell_{11}^2+\ell_{10}\ell_{21}\ell_{11}+\ell_{10}^2\ell_{22}=0.$$ 
Plugging the expressions of $\ell_{ij}$ for $0\le i,j\le2$ into the above equation, we have 
\begin{equation*}
\theta_1^3+\theta_1^2\theta_4+\theta_1\theta_2\overline{\theta}_2+\theta_2^2\theta_3+\overline{\theta}_2^2\overline{\theta}_3 = 0,
\end{equation*}
i.e.,
\begin{equation}
\label{2_theta}
\theta_4 = \frac{\theta_1^3+\theta_1\theta_2\overline{\theta}_2+\theta_2^2\theta_3+\overline{\theta}_2^2\overline{\theta}_3 }{\theta_1^2}. 
\end{equation}
In addition, from $\tr_1^m\left( \frac{\ell_{20}\ell_{22}}{\ell_{21}^2} \right)=1$, we have 
\begin{equation}
\label{tr_1}
\tr_1^m\left( \frac{(\theta_1+\theta_3+\overline{\theta}_3)(\theta_4 + \omega \theta_3 + \omega^2 \overline{\theta}_3 + \omega\theta_2 + \omega^2 \overline{\theta}_2)  }{\theta_1^2+\theta_2^2+\overline{\theta}_2^2} \right) =1.
\end{equation}
Using the following relation,
$$\tr_1^m\left(\frac{\omega\theta_3^2+\omega^2\overline{\theta}_3^2}{\theta_1^2+\theta_2^2+\overline{\theta}_2^2}\right) = \tr_1^m \left( \frac{(\omega^2\theta_3+\omega\overline{\theta}_3)(\theta_1+\theta_2+\overline{\theta}_2)}{\theta_1^2+\theta_2^2+\overline{\theta}_2^2} \right), $$
Eq. (\ref{tr_1}) can be simplified as 
\begin{equation}
\label{tr_2}
\tr_1^m\left( \frac{ \omega\theta_1\theta_2 + \omega^2\theta_1\overline{\theta}_2+\theta_1\theta_3 + \theta_1\overline{\theta}_3 + \theta_2\theta_3 + \overline{\theta}_2\overline{\theta}_3 + \theta_3\overline{\theta}_3 + \theta_4 (\theta_1+\theta_3+\overline{\theta}_3)   }{\theta_1^2+\theta_2^2+\overline{\theta}_2^2}   \right) =1
\end{equation}
Next plugging Eq. (\ref{2_theta}) into Eq. (\ref{tr_2}) and simplifying, we obtain
\begin{equation}
\label{tr_3}
\tr_1^m\left( \frac{\Delta}{\theta_1^2\left( \theta_1+\theta_2+\overline{\theta}_2  \right)^2} \right) =1,
\end{equation}
where
\begin{eqnarray*}
	\Delta &=& \theta_1^4+\omega \theta_1^3\theta_2  + \omega^2\theta_1^3\overline{\theta}_2 + \theta_1^2\theta_2\theta_3 + \theta_1^2\theta_2\overline{\theta}_2 + \theta_1^2\theta_3\overline{\theta}_3 + \theta_1^2\overline{\theta}_2\overline{\theta}_3+\theta_1\theta_2^2\theta_3  \\
	&&+ \theta_1\theta_2\overline{\theta}_2\theta_3 +\theta_1\theta_2\overline{\theta}_2\overline{\theta}_3 + \theta_1\overline{\theta}_2^2\overline{\theta}_3+\theta_2^2\theta_3^2+\overline{\theta}_2^2\overline{\theta}_3^2 +\overline{\theta}_2^2\theta_3\overline{\theta}_3+\theta_2^2\theta_3\overline{\theta}_3.
\end{eqnarray*}
Also, using the following relation, 
$$\tr_1^m\left(  \frac{\theta_2^2\theta_3^2+\overline{\theta}_2^2\overline{\theta}_3^2}{\theta_1^2\left( \theta_1+\theta_2+\overline{\theta}_2  \right)^2} \right)=\tr_1^m\left( \frac{\left(\theta_2\theta_3+\overline{\theta}_2\overline{\theta}_3 \right) \theta_1\left( \theta_1+\theta_2+\overline{\theta}_2  \right)}{\theta_1^2\left( \theta_1+\theta_2+\overline{\theta}_2  \right)^2} \right),$$
Eq. (\ref{tr_3}) becomes
\begin{eqnarray*}
	1 &=& \tr_1^m\left( \frac{ \theta_1^4+\omega \theta_1^3\theta_2 +\omega^2\theta_1^3\overline{\theta}_2+\theta_1^2\theta_2\overline{\theta}_2+\theta_1^2\theta_3\overline{\theta}_3+\overline{\theta}_2^2\theta_3\overline{\theta}_3+\theta_2^2\theta_3\overline{\theta}_3  }{\theta_1^2\left( \theta_1+\theta_2+\overline{\theta}_2  \right)^2} \right)\\
	&=& \tr_1^m\left(\frac{\theta_1^2+\omega\theta_1\theta_2+\omega^2\theta_1\overline{\theta}_2+\theta_2\overline{\theta}_2}{\left( \theta_1+\theta_2+\overline{\theta}_2  \right)^2}\right) + \tr_1^m \left( \frac{\theta_3\overline{\theta}_3}{\theta_1^2} \right).
\end{eqnarray*}
In addition, 
\begin{eqnarray*}
	&&\tr_1^m\left(\frac{\theta_1^2+\omega\theta_1\theta_2+\omega^2\theta_1\overline{\theta}_2+\theta_2\overline{\theta}_2}{\left( \theta_1+\theta_2+\overline{\theta}_2  \right)^2}\right)\\
	&=& \tr_1^m\left(\frac{\omega^2\theta_1^2 +\overline{\theta}_2^2 + \omega\theta_1(\theta_1+\theta_2+\overline{\theta}_2)+\overline{\theta}_2(\theta_1+\theta_2+\overline{\theta}_2)}{\left( \theta_1+\theta_2+\overline{\theta}_2  \right)^2}\right)\\
	&=& \tr_1^m\left( \frac{\omega^2\theta_1^2+\overline{\theta}_2^2}{\left( \theta_1+\theta_2+\overline{\theta}_2  \right)^2} + \frac{\omega^2\theta_1+\overline{\theta}_2}{\theta_1+\theta_2+\overline{\theta}_2  } \right)\\
	&=&1
\end{eqnarray*}
since $\omega^2\theta_1+\overline{\theta}_2 \in \gf_{2^{n}} \backslash \gf_{2^{m}}$ (if $\omega^2\theta_1+\overline{\theta}_2 \in \gf_{2^{m}}$, $\theta_1+\theta_2+\overline{\theta}_2=0$, which contradicts our assumption). Thus we have 
\begin{equation}
\label{tr_4}
\tr_1^m \left( \frac{\theta_3\overline{\theta}_3}{\theta_1^2} \right) = 0.
\end{equation}

Moreover, from Lemma \ref{TLZ}, we get $$\theta_2\overline{\theta}_2+\theta_3\overline{\theta}_3=\theta_4\left(\theta_1+\theta_4\right).$$
Plugging Eq. (\ref{2_theta}) into the above equation, we obtain
\begin{equation}
\label{2_theta_1}
\theta_1^4\theta_3\overline{\theta}_3+\theta_1^3  \theta_2^2\theta_3+ \theta_1^3 \overline{\theta}_2^2\overline{\theta}_3   + \theta_1^2\theta_2^2\overline{\theta}_2^2+ \theta_2^4\theta_3^2+\overline{\theta}_2^4\overline{\theta}_3^2 = 0.
\end{equation}
Assume $\theta_2^2 = \theta_1\overline{\theta}_3 + \epsilon,$ where $\epsilon\neq0$. Then $\overline{\theta}_2^2 =\theta_1\theta_3+\overline{\epsilon}$. Plugging the expressions of $\theta_2$ and $\overline{\theta}_2$ into Eq. (\ref{2_theta_1}) directly and simplifying, we get 
\begin{equation}
\label{2_theta_2}
\theta_1^2 = \frac{\epsilon}{\overline{\epsilon}} \theta_3^2+ \frac{\overline{\epsilon}}{\epsilon}\overline{\theta}_3^2.
\end{equation} 
Plugging Eq. (\ref{2_theta_2}) into Eq. (\ref{tr_4}), we obtain
\begin{eqnarray*}
	0&=& \tr_1^m\left( \frac{\epsilon\theta_3\overline{\epsilon}\overline{\theta}_3}{\epsilon^2\theta_3^2+\overline{\epsilon}^2\overline{\theta}_3^2}  \right) \\
	&=& \tr_1^m \left( \frac{t}{ t^2 + 1}  \right) \\
	&=& \tr_1^m \left(  \frac{1}{ t+ 1}  + \frac{1}{t^2 + 1}  \right) \\
	&=& 1, 
\end{eqnarray*}
where $t = \frac{\overline{\epsilon}\overline{\theta}_3}{\epsilon\theta_3}\in\gf_{2^{n}}\backslash\gf_{2^{m}}$. Conflict! Therefore, $\epsilon=0$, i.e., $$\theta_2^2 = \theta_1\overline{\theta}_3.$$ 

Furthermore, when $\theta_2^2 = \theta_1\overline{\theta}_3,$  from Eq. (\ref{2_theta}), we get $\theta_4^2=\theta_1^2+\theta_3\overline{\theta}_3$ directly. In addition,  we have $\tr_1^m\left( \frac{\theta_4}{\theta_1} \right) =1$ according to $\tr_1^m\left( \frac{\ell_{20}\ell_{22}}{\ell_{21}^2} \right)=1$, see Subcase 1.1. Moreover, $\tr_1^m\left(\frac{\ell_{00}\ell_{22}}{\ell_{11}^2}\right) =1$ is also equivalent to $\tr_1^m\left( \frac{\theta_4}{\theta_1} \right) =1$ since
\begin{eqnarray*}
	& & \tr_1^m\left( \frac{(\theta_1+\theta_3+\overline{\theta}_3)(\theta_1+\omega^2\theta_3+\omega\overline{\theta}_3)}{\theta_1^2}  \right) \\
	&=& \tr_1^m\left( \frac{\theta_1^2+\omega\theta_1\theta_3+\omega^2\theta_1\overline{\theta}_3+\theta_3\overline{\theta}_3}{\theta_1^2} + \frac{\omega^2\theta_3^2+\omega\overline{\theta}_3^2}{\theta_1^2}  \right) \\
	&=& \tr_1^m\left( \frac{\theta_1^2+\omega\theta_1\theta_3+\omega^2\theta_1\overline{\theta}_3+\theta_3\overline{\theta}_3}{\theta_1^2} +  \frac{\theta_1(\omega\theta_3+\omega\overline{\theta}_3)}{\theta_1^2} \right) \\
	&=& \tr_1^m\left(\frac{\theta_1^2+\theta_3\overline{\theta}_3}{\theta_1^2}\right) = \tr_1^m\left( \frac{\theta_4}{\theta_1} \right) =1.
\end{eqnarray*}

Thus  there exist some  $a,b \in\gf_{2^{n}}\backslash\gf_{2^{m}}$, such that $L(x,y)=\ell_{22}(xy+ax+ay+b)(xy+\overline{a}x+\overline{a}y+\overline{b})$ if and only if 
$$ \theta_1\neq0, \theta_1\neq\theta_3+\overline{\theta}_3, \theta_1\neq\theta_2+\overline{\theta}_2, \theta_2^2 = \theta_1\overline{\theta}_3, \theta_4^2=\theta_1^2+\theta_3\overline{\theta}_3, \tr_1^m\left( \frac{\theta_4}{\theta_1} \right) =1. $$

When $\ell_{11}=\theta_1\neq0$ and $\ell_{21} = \theta_1+\theta_2+\overline{\theta}_2 =0,$ we have $\theta_2\in \gf_{2^{n}}\backslash \gf_{2^{m}}$, $a=\overline{a}$ and $a^2=\frac{\ell_{20}}{\ell_{22}}$.  From (\ref{2_l_3}) and (\ref{2_l_5}), we have $\tr_1^m\left( \frac{\ell_{00}\ell_{22}}{\ell_{11}^2} \right)=1$, i.e.,
$$\tr_1^m\left( \frac{\theta_1^2+\omega \theta_1\theta_3+\omega^2\theta_1\overline{\theta}_3+\omega^2\theta_3^2+\theta_3\overline{\theta}_3+\omega\overline{\theta}_3^2}{\theta_1^2} \right)=1.$$
Moreover, it is clear that 
\begin{eqnarray*}
	&&\tr_1^m\left( \frac{\omega \theta_1\theta_3+\omega^2\theta_1\overline{\theta}_3+\omega^2\theta_3^2+\omega\overline{\theta}_3^2}{\theta_1^2} \right) \\
	&=&\tr_1^m\left( \frac{\omega\theta_3+\omega^2\overline{\theta}_3}{\theta_1} +  \frac{(\omega\theta_3+\omega^2\overline{\theta}_3)^2}{\theta_1^2}  \right) \\
	&=& 0,
\end{eqnarray*}
and thus $$\tr_1^m\left(\frac{\theta_3\overline{\theta}_3}{\theta_2^2+\overline{\theta}_2^2}\right) = \tr_1^m\left(\frac{\theta_3\overline{\theta}_3}{\theta_1^2}\right)=0.$$
In addition, from Lemma \ref{TLZ}, we get $$\theta_2\overline{\theta}_2+\theta_3\overline{\theta}_3=\theta_4\left(\theta_1+\theta_4\right).$$ Plugging $\theta_1=\theta_2+\overline{\theta}_2$ into the above equation, we have 
$$\theta_3\overline{\theta}_3 = \theta_2\theta_4+\theta_2\overline{\theta}_2+\theta_4^2+\theta_4\overline{\theta}_2.$$
Therefore, 
\begin{eqnarray*}
	& &\tr_1^m\left(\frac{\theta_3\overline{\theta}_3}{\theta_2^2+\overline{\theta}_2^2}\right) \\
	&=& \tr_1^m\left( \frac{\theta_4(\theta_2+\overline{\theta}_2)+\theta_4^2+\theta_2\overline{\theta}_2}{\theta_2^2+\overline{\theta}_2^2}  \right) \\
	&=& \tr_1^m\left(\frac{\theta_2\overline{\theta}_2}{\theta_2^2+\overline{\theta}_2^2}\right) \\
	&=& \tr_1^m\left(\frac{1}{t+1}+\frac{1}{t^2+1}\right) =1, 
\end{eqnarray*}
where $t=\frac{\overline{\theta}_2}{\theta_2}\in\gf_{2^{n}}\backslash\gf_{2^{m}}$. Conflict! Hence, there does not exist any $a\in\gf_{2^{m}}, b \in \gf_{2^{n}}$ such that $L(x,y)=\ell_{22}(xy+ax+ay+b)(xy+\overline{a}x+\overline{a}y+\overline{b})$.

When $\ell_{21}=\theta_1+\theta_2+\overline{\theta}_2\neq0$ and $\ell_{11}=\theta_1=0$, $b=\overline{b}$ and $b^2=\frac{\ell_{00}}{\ell_{22}}$. From (\ref{2_l_1}) and (\ref{2_l_2}), we have $\tr_1^m\left( \frac{\ell_{20}\ell_{22}}{\ell_{21}^2} \right)=1$. Moreover, computing $ (\ref{2_l_4})^2 \times \ell_{22} + (\ref{2_l_5})\times(\ref{2_l_1})^2$, we get 
$$ \ell_{10}^2\ell_{22}=\ell_{00}\ell_{21}^2, $$
i.e.,
$$ \theta_2^2\theta_3 = \overline{\theta}_2^2\overline{\theta}_3. $$
Thus $\theta_2^2\theta_3\in\gf_{2^{m}}$, denoted by $\epsilon\in\gf_{2^{m}}$. 
In addition, from $\tr_1^m\left( \frac{\ell_{20}\ell_{22}}{\ell_{21}^2} \right)=1$, we have 
\begin{equation}
\label{tr_5} 
\tr_1^m \left( \frac{\omega \theta_2\theta_3+ \omega \theta_2\overline{\theta}_3 + \theta_3\theta_4+\omega^2\overline{\theta}_2\theta_3+\theta_3\overline{\theta}_3+\overline{\theta}_3\theta_4+\omega^2\overline{\theta}_2\overline{\theta}_3+\omega\theta_3^2+ \omega^2\overline{\theta}_3^2}{\theta_2^2+\overline{\theta}_2^2}  \right) =1.
\end{equation}
Also, thanks to 
$$ \tr_1^m\left( \frac{\omega\theta_3^2+ \omega^2\overline{\theta}_3^2}{\theta_2^2+\overline{\theta}_2^2} \right) =\tr_1^m\left( \frac{ \left( \omega^2\theta_3+ \omega\overline{\theta}_3 \right)\left( \theta_2+\overline{\theta}_2 \right) }{\theta_2^2+\overline{\theta}_2^2} \right), $$
together with Eq. (\ref{tr_5}), we obtain 
\begin{equation}
\label{tr_6}
\tr_1^m\left( \frac{\theta_2\theta_3+\theta_3\theta_4+\theta_3\overline{\theta}_3+\overline{\theta}_3\theta_4+\overline{\theta}_2\overline{\theta}_3}{\theta_2^2+\overline{\theta}_2^2} \right)=1.
\end{equation}
Plugging $\theta_3=\frac{\epsilon}{\theta_2^2}$ and $\overline{\theta}_3=\frac{\epsilon}{\overline{\theta}_2^2}$ into Eq. (\ref{tr_6}) and simplifying, we get 
\begin{eqnarray*}
	&&\tr_1^m\left( \frac{\theta_2\theta_3+\theta_3\theta_4+\theta_3\overline{\theta}_3+\overline{\theta}_3\theta_4+\overline{\theta}_2\overline{\theta}_3}{\theta_2^2+\overline{\theta}_2^2}  \right) \\
	&=& \tr_1^m\left( \frac{\epsilon/\theta_2 + \theta_4\epsilon (1/\theta_2^2+1/\overline{\theta}_2^2) + \epsilon^2/(\theta_2^2\overline{\theta}_2^2) + \epsilon/\overline{\theta}_2 }{\theta_2^2+\overline{\theta}_2^2}   \right) \\
	&=&\tr_1^m\left( \frac{\epsilon\theta_2\overline{\theta}_2(\theta_2+\overline{\theta}_2)+\epsilon^2+\theta_4\epsilon(\theta_2^2+\overline{\theta}_2^2)}{\theta_2^2\overline{\theta}_2^2\left(\theta_2^2+\overline{\theta}_2^2\right)}  \right) \\
	&=& \tr_1^m\left(\frac{\theta_4\epsilon}{\theta_2^2\overline{\theta}_2^2}\right)=1.
\end{eqnarray*}
In addition, from Lemma \ref{TLZ}, we get $$\theta_2\overline{\theta}_2+\theta_3\overline{\theta}_3=\theta_4\left(\theta_1+\theta_4\right).$$ Plugging $\theta_1=0$ into the above equation, we have 
$$\theta_4^2=\theta_2\overline{\theta}_2+\theta_3\overline{\theta}_3.$$
Furthermore,
\begin{eqnarray*}
	\tr_1^m\left(\frac{\theta_4\epsilon}{\theta_2^2\overline{\theta}_2^2}\right) &=& \tr_1^m\left(\frac{\theta_4^2\epsilon^2}{\theta_2^4\overline{\theta}_2^4}\right) \\
	&=& \tr_1^m\left( \frac{\left(\theta_2\overline{\theta}_2+\theta_3\overline{\theta}_3\right)\epsilon^2}{\theta_2^4\overline{\theta}_2^4}  \right) \\
	&=& \tr_1^m\left( \frac{\left(\theta_2\overline{\theta}_2+\epsilon^2/(\theta_2^2\overline{\theta}_2^2)\right)\epsilon^2}{\theta_2^4\overline{\theta}_2^4}  \right) \\
	&=& \tr_1^m\left( \frac{\epsilon^4}{\theta_2^6\overline{\theta}_2^6} + \frac{\epsilon^2\theta_2^3\overline{\theta}_2^3}{\theta_2^6\overline{\theta}_2^6}  \right) =0,
\end{eqnarray*}
which is a contradiction! Thus, there does not exist any $a\in\gf_{2^n}, b \in \gf_{2^{m}}$ such that $L(x,y)=\ell_{22}(xy+ax+ay+b)(xy+\overline{a}x+\overline{a}y+\overline{b})$.

All in all, there exist some $a,b\in\gf_{2^{n}}$ with $a,b$ not in $\gf_{2^{m}}$ at the same time such that $L(x,y)=\ell_{22}(xy+ax+ay+b)(xy+\overline{a}x+\overline{a}y+\overline{b})$ if and only if 
$$ \theta_1\neq0, \theta_1\neq\theta_3+\overline{\theta}_3, \theta_1\neq\theta_2+\overline{\theta}_2, \theta_2^2 = \theta_1\overline{\theta}_3, \theta_4^2=\theta_1^2+\theta_3\overline{\theta}_3, \tr_1^m\left( \frac{\theta_4}{\theta_1} \right) =1. $$

As for (iii), there exist some $a\in\overline{\gf}_{2^m}\backslash\gf_{2^{m}}$ and $b\in\gf_{2^{m}}$ such that
\begin{eqnarray*}
	L &=&\ell_{22}(xy+ax+\overline{a}y+b)(xy+\overline{a}x+{a}y+b) \\
	&=& \ell_{22}\left( x^2y^2+(\overline{a}+a)x^2y+ (\overline{a}+a)xy^2 + a\overline{a} x^2 + \right. \\
	&&\left.(a^2+\overline{a}^2) xy + a\overline{a} y^2 +(ab+\overline{a}b)x+ (ab+\overline{a}b)y+ b^2 \right).
\end{eqnarray*}
Comparing the coefficients of the above expression and (\ref{L(x,y)}), we have 
\begin{subequations} 	
	\renewcommand\theequation{\theparentequation.\arabic{equation}}     
	\begin{empheq}[left={\empheqlbrace\,}]{align}
	~~	\ell_{21} &=  (a+\overline{a})\ell_{22}  \label{3_l_1} \\ 
	~~	\ell_{20} &= a\overline{a}\ell_{22}     \label{3_l_2} \\
	~~	\ell_{11} &= (a^2+\overline{a}^2)\ell_{22} \label{3_l_3} \\
	~~	\ell_{10} &= (ab+ \overline{a} b)\ell_{22} \label{3_l_4} \\
	~~	\ell_{00} &= b^2\ell_{22}. \label{3_l_5}
	\end{empheq}
\end{subequations}
Computing $(\ref{3_l_1})^2+(\ref{3_l_3})\times \ell_2$ and $ (\ref{3_l_4})^2 + (\ref{3_l_3})\times (\ref{3_l_5}) $, we obtain 
\begin{equation*}
\ell_{21}^2+\ell_{11}\ell_{22}=0 ~\text{and}~ \ell_{11}\ell_{00} +\ell_{10}^2=0,
\end{equation*}
i.e.,
\begin{equation}
\label{3_l_6}
\theta_1\theta_3+\theta_1\overline{\theta}_3+\theta_2^2+\overline{\theta}_2^2=0
\end{equation}
and 
\begin{equation}
\label{3_l_7}
\omega^2\theta_1\theta_3+\omega\theta_1\overline{\theta}_3+\omega\theta_2^2+\omega^2\overline{\theta}_2^2=0.
\end{equation}
Computing $(\ref{3_l_6})+\omega\times (\ref{3_l_7})$, we have $ \omega^2 \left(\theta_1\overline{\theta}_3+\theta_2^2\right) =0 $ and thus $\theta_2^2=\theta_1\overline{\theta}_3.$ In addition, since $a+\overline{a}\neq0$ and $\ell_{11} = (a^2+\overline{a}^2)\ell_{22}$, we know $\theta_1= \ell_{11}\neq0$. Furthermore, when $\theta_1\neq0$ and $\theta_2^2=\theta_1\overline{\theta}_3$, $a, \overline{a}$ are solutions in $\gf_{2^{n}}$ of 
$$x^2+\frac{\ell_{21}}{\ell_{22}}x+\frac{\ell_{20}}{\ell_{22}}=0$$
and thus $\tr_1^m\left(\frac{\theta_4}{\theta_1}\right)=1$, see Subcase 1.1. Hence, there exist some $a\in\gf_{2^{n}}, b\in\gf_{2^m}$ such that  $L =\ell_{22}(xy+ax+\overline{a}y+b)(xy+\overline{a}x+{a}y+b)$ if and only if $$\theta_1\neq0,  \theta_1\neq\theta_3+\overline{\theta}_3, \theta_2^2=\theta_1\overline{\theta}_3, \tr_1^m\left(\frac{\theta_4}{\theta_1}\right)=1.$$

{ \bfseries Case 2: $\ell_{22}=0$.} Namely,  $\theta_1=\theta_3+\overline{\theta}_3.$

In the case, if $\ell_{21}\neq0$, then $\deg L =3$. We also use the important property that the morphisms $(x,y)\to (y,x)$ and $(x,y) \to (\overline{x},\overline{y})$ fix $L=0$. If
 $L$ has a component $L_1$ {not defined over $\gf_{2^{m}}$} with degree $1$. Then $\overline{L}_1\neq L$   with degree $1$ is also a component of $L$,  where $\overline{L}_1$ denotes the polynomial from raising all the coefficients of $L_1$ into their $2^m$-th power. Thus if $L$ can split completely into absolute irreducible components not defined over $\gf_{2^{m}}$, the only possibility is $(1,1,1)$. Namely, there exist some $a,b\in\overline{\gf}_{2^m}\backslash\gf_{2^{m}}$ such that $L(x,y)=\ell_{21}(x+y+a)(x+b)(y+b).$ However, since $L(\overline{x}_0,\overline{y}_0)=0$ if $L(x_0,y_0)=0$, we have $a,b\in\gf_{2^m}.$ Conflict! Therefore, $\ell_{21}=0$. 

When $\ell_{22}=\ell_{21}=0$, $\theta_1=\theta_2+\overline{\theta}_2=\theta_3+\overline{\theta}_3$ and thus $\theta_2+\theta_3\in\gf_{2^{m}}$, denoted by $\gamma\in\gf_{2^{m}}$. Moreover, we have 
\begin{equation}
\label{L_2}L(x,y) = \ell_{20}x^2+\ell_{11}xy+\ell_{02}y^2+\ell_{10}x+\ell_{01}y+\ell_{00},
\end{equation}
where 
\begin{subequations} 	
	\renewcommand\theequation{\theparentequation.\arabic{equation}}     
	\begin{empheq}[left={\empheqlbrace\,}]{align}
	~~	\ell_{20} &= \ell_{02} = \theta_4+\gamma    \label{2_L_1} \\
	~~	\ell_{11} &=  \theta_2+\overline{\theta}_2 \label{2_L_2} \\
	~~	\ell_{10} &= \ell_{01}= \omega \theta_2 + \omega^2\overline{\theta}_2 \label{2_L_3} \\
	~~	\ell_{00} &= \omega \theta_2 + \omega^2\overline{\theta}_2+ \gamma, \label{2_L_4}
	\end{empheq}
\end{subequations}
If $\ell_{20}=0$, i.e., $\theta_4=\gamma$, to make sure that $L(x,y)$ does not have absolute irreducible components defined over $\gf_{2^{m}}$ different from $y=x$, we have $\ell_{11}=\ell_{00}=0$. Namely, $\theta_2\in\gf_{2^{m}}$ and $\gamma=\theta_2$. Furthermore, $\theta_3=0$ and $\theta_1=0$.  Considering the expressions of $\theta_i$, we have 
\begin{subequations} 	
	\renewcommand\theequation{\theparentequation.\arabic{equation}}     
	\begin{empheq}[left={\empheqlbrace\,}]{align}
	~~	 &1 + a_1^2+a_2\overline{a}_2+a_3\overline{a}_3=0     \label{a_1} \\
	~~	 & \overline{a}_2+a_1\overline{a}_3=0 \label{a_2} \\
	~~	 & a_1+\overline{a}_2a_3=a_1+a_2\overline{a}_3 \label{a_3} \\
	~~	 & a_1+\overline{a}_2a_3=a_1^2+a_2\overline{a}_2. \label{a_4}
	\end{empheq}
\end{subequations}
Plugging (\ref{a_4}) into (\ref{a_1}), we have 
$$1+a_1+\overline{a}_2a_3+a_3\overline{a}_3=0,$$
i.e.,
$$(a_1+1)(a_3\overline{a}_3+1)=0.$$
Thus $a_1=1$ or $a_3\overline{a}_3=1$. If $a_1=1$, we have $a_2=a_3$ from (\ref{a_2}), which means $1+a_1+a_2+a_3=0$, which is a contradiction. If $a_3\overline{a}_3=1$, then $\theta_2=\theta_4=a_1^2+a_2\overline{a}_2=0$, which means $L=0$, which is also impossible.

Therefore, $\ell_{20}\neq0$. Suppose that there exist some $a,b\in\gf_{2^{n}}$ such that $$L=\ell_{20} (x+ay+b)(x+\overline{a}y+\overline{b}). $$ After comparing the coefficients of the above expression and (\ref{L_2}), we have $a=1$ or $a=\frac{b}{\overline{b}}$.

If $a=1$, then $L=\ell_{20}\left( x^2+y^2+(b+\overline{b})x + (b+\overline{b})y + b\overline{b} \right)$ and thus $\ell_{11}=\theta_2+\overline{\theta}_2=0$, i.e., $\theta_2\in\gf_{2^{m}}$. Furthermore, $\theta_1=0$ and $\theta_3=\gamma+\theta_2\in\gf_{2^{m}}$. In addition, from Lemma \ref{TLZ}, we get $\theta_2\overline{\theta}_2+\theta_3\overline{\theta}_3=\theta_4\left(\theta_1+\theta_4\right)$ and thus $\theta_4=\theta_2+\theta_3=\gamma$, which means $\ell_{20}=0$. Conflict.

If $a=\frac{b}{\overline{b}}$,then  $L=\ell_{20}\left( x^2+y^2+ (\frac{b}{\overline{b}}+\frac{\overline{b}}{{b}})xy +(b+\overline{b})x + (b+\overline{b})y + b\overline{b} \right)$ and thus 
\begin{subequations} 	
	\renewcommand\theequation{\theparentequation.\arabic{equation}}     
	\begin{empheq}[left={\empheqlbrace\,}]{align}
	~~ &(\frac{b}{\overline{b}}+\frac{\overline{b}}{{b}})\ell_{20} =	\ell_{11} =  \theta_2+\overline{\theta}_2 \label{3_L_1} \\
	~~& (b+\overline{b}) \ell_{20} =	\ell_{10} = \omega \theta_2 + \omega^2\overline{\theta}_2 \label{3_L_2} \\
	~~&	b\overline{b} \ell_{20} = \ell_{00} = \omega \theta_2 + \omega^2\overline{\theta}_2+ \gamma, \label{3_L_3}.
	\end{empheq}
\end{subequations}
Computing $(\ref{3_L_2})^2/(\ref{3_L_3})+(\ref{3_L_1})$, we obtain $\ell_{10}^2=\ell_{00}\ell_{11}$, i.e., 
$$\gamma=\frac{\theta_2^2+\overline{\theta}_2^2+\theta_2\overline{\theta}_2}{\theta_2+\overline{\theta}_2}.$$
In addition, from Lemma \ref{TLZ}, we get $\theta_2\overline{\theta}_2+\theta_3\overline{\theta}_3=\theta_4\left(\theta_1+\theta_4\right)$. Plugging $\theta_3=\theta_2+\gamma$ and $\theta_1=\theta_2+\overline{\theta}_2$ into the above equation and simplifying, we have 
$$(\theta_4+\gamma)(\theta_4+\gamma+\theta_2+\overline{\theta}_2)=0.$$
Hence, $\theta_4=\gamma+\theta_2+\overline{\theta}_2 $ since $\ell_{20}=\theta_4+\gamma\neq0$. Moreover, from (\ref{3_L_2}) and (\ref{3_L_3}),  we know that $b\in\gf_{2^{n}}\backslash\gf_{2^{m}}$ if and only if $\tr_1^m\left( \frac{\ell_{00}\ell_{20}}{\ell_{10}^2} \right)=1$, which holds under our assumption. In fact,  
\begin{eqnarray*}
	\tr_1^m\left( \frac{\ell_{00}\ell_{20}}{\ell_{10}^2} \right) &=& \tr_1^m\left( \frac{(\omega\theta_2+\omega^2\overline{\theta}^2+\gamma)(\theta_2+\overline{\theta}_2)}{\omega^2\theta_2^2+\omega\overline{\theta}_2^2}  \right) \\
	&=& \tr_1^m\left( \frac{\omega\theta_2^2+\omega^2\overline{\theta}_2^2+\theta_2\overline{\theta}_2+\gamma(\theta_2+\overline{\theta}_2)}{\omega^2\theta_2^2+\omega\overline{\theta}_2^2} \right)\\
	&=& \tr_1^m(1)=1
\end{eqnarray*}
since $\gamma=\frac{\theta_2^2+\overline{\theta}_2^2+\theta_2\overline{\theta}_2}{\theta_2+\overline{\theta}_2}$ and $m$ is odd. Therefore,  there exist some $ a, b\in\gf_{2^n}$ such that  $L=\ell_{20} (x+ay+b)(x+\overline{a}y+\overline{b})$ if and only if 
$$ \theta_1=\theta_2+\overline{\theta}_2\neq0, \theta_3=\theta_2+\gamma, \theta_4=\gamma+\theta_2+\overline{\theta}_2 ~~\text{and}~~ \gamma=\frac{\theta_2^2+\overline{\theta}_2^2+\theta_2\overline{\theta}_2}{\theta_2+\overline{\theta}_2}. $$

\bibliographystyle{plain}

\bibliography{ref}

\end{document}